 \documentclass[structabstract,letter]{aa}   
\usepackage{graphicx}
\usepackage{txfonts}
\usepackage{epsf}  
\usepackage{amssymb}  
\usepackage{lscape} 
\usepackage{natbib}
\bibpunct{(}{)}{;}{a}{}{,} 
\begin{document}
%
%
 \newcommand{\betacrbbold}{\mbox{\boldmath $\beta$}~CrB} 

 \newcommand{\betacrb}{$\beta$~CrB}
 \newcommand{\tenaql}{10~Aql}
 \newcommand{\gammaequ}{$\gamma$~Equ}

 \newcommand{\thetald}{$\theta_{\rm LD}$}
 \newcommand{\fbol}{$f_{\rm bol}$}
 \newcommand{\acir}{$\alpha$~Cir}
 \newcommand{\ie}{i.e.}
 \newcommand{\eg}{e.g.}
 \newcommand{\cf}{cf.}
 \newcommand{\kms}{km\,s$^{-1}$}
 \newcommand{\teff}{$T_{\rm eff}$}
 \newcommand{\logg}{$\log g$}
 \newcommand{\feh}{[Fe/H]}
 \newcommand{\msun}{${\rm M}_{\odot}$}
 \newcommand{\percent}{\,{\%}}
\newcommand{\kepler}{{\em Kepler}}

 \newcommand{\dss}{$\delta$~Scuti star}
 \newcommand{\ngc}{NGC~6866}

\newcommand{\vsini}{$v \sin i$}
\newcommand{\feone}{Fe\,{\sc I}}
\newcommand{\fetwo}{Fe\,{\sc II}}

%


%
%
\title{Atmospheric parameters of red giants in the {\em Kepler} field.
\thanks{Based on observations made with the Nordic Optical Telescope, operated
on the island of La Palma jointly by Denmark, Finland, Iceland,
Norway, and Sweden, in the Spanish Observatorio del Roque de los
Muchachos of the Instituto de Astrofisica de Canarias.}}
\titlerunning{Atmospheric parameters of red giants}
\authorrunning{H. Bruntt et al.}
\author{
H.~Bruntt\inst{1} 
\and
S.~Frandsen\inst{1}
\and
A. O.~Thygesen\inst{1,2}
} 
\offprints{S.~Frandsen}
\mail{srf@phys.au.dk}
\institute{Department of Physics and Astronomy, Aarhus University, DK-8000 Aarhus C, Denmark.
\and Nordic Optical Telescope, Apartado 474, E-38700 Santa Cruz de La Palma, Santa Cruz de Tenerife, Spain.}
\date{Received ??-?? 2010 ; Accepted ??-?? ???? }
\abstract
{Accurate fundamental parameters of stars are mandatory for the asteroseismic investigation of the \kepler\ mission to succeed.} 
{We will determine the atmospheric parameters for a sample of 6 well-studied bright K~giants to confirm that our method produces reliable results. We then apply the same method to 14 K~giants that are targets for the \kepler\ mission.}
{We have used high-resolution, high signal-to-noise spectra from the FIES spectrograph on the Nordic Optical Telescope. We used the iterative spectral synthesis method VWA to derive the fundamental parameters from carefully selected high-quality iron lines and pressure-sensitive Calcium lines.}
{We find good agreement with parameters from the literature for the 6 bright giants.
We compared the spectroscopic values with parameters based on photometric indices in the Kepler Input Catalogue (KIC).
We identify serious problems with the KIC values for \feh\ and find a large RMS scatter of 0.5 dex.
The \logg\ values in KIC agree reasonably well with the spectroscopic values 
with a scatter of 0.25 dex, when excluding two low-metallicity giants.
The \teff\,s from VWA and KIC agree well with a scatter of about 85\,K.
We also find good agreement with \logg\ and \teff\ derived from 
asteroseismic analyses for seven \kepler\ giant targets.}
{We have determined accurate fundamental parameters of 14 giants
using spectroscopic data. The large discrepancies between photometric and 
spectroscopic values of \feh\ emphasize the need for further 
detailed spectroscopic follow-up of the \kepler\ targets. 
This is mandatory to be able to produce reliable constraints 
for detailed asteroseismic analyses and for the interpretation 
of possible exo-planet candidates found around giant stars.}
\maketitle
%
%
%
\section{Introduction \label{sec:intro}}

During 2009 the space missions CoRoT and \kepler\ have generated a high level of activity in
the groups specialized in the asteroseismic analysis of photometric time series of stars. In both cases
the very long, continuous observing and the very low noise data has opened up a completely new
world of possibilities for the asteroseismic investigation of stellar interiors.

In particular, the seismic investigation of K~giants has taken 
a huge leap forward. 
The results from a time series analysis of 150 days of measurements obtained by
the CoRoT space telescope increased the number of known pulsating giants from a handful to
nearly 800 \citep{deridder}. The \kepler\ mission is observing the flux continuously of
thousands of stars for at least 3 years and has increased both the number, 
the range in luminosity, and
the length of the time series compared to CoRoT. 
The high-precision light curves from \kepler\ 
constitute important data for detailed
asteroseismic investigations of red giants due to the long temporal coverage 
and low noise levels of the observations. 
This has extended the range of giants with detected oscillations
to lower luminosities \citep{bedding,stello3,mosser}. 

Before we can hope to make a successful analysis of individual
red giant stars observed by \kepler\, we need to measure accurate atmospheric parameters
as discussed by \citet{brown}, \citet{creevey1}, and \citet{creevey2}.

We have started the observations of about 100 
\kepler\ red giant stars with the FIbre-fed Echelle Spectrograph (FIES) spectrograph 
at the Nordic Optical Telescope (NOT).  
From the spectral analysis we can determine accurate
atmospheric parameters, which are essential for constraining the stellar models when comparing
asteroseismic observations and theory. We concentrate in particular on old, metal poor stars,
which are important for the understanding the early history of the Galaxy. We will also obtain
important insight about the observed variation of the pulsational behaviour with metallicity.

At present only photometric determinations of the metallicity are available, based on the Kepler Input
Catalogue \citep{kic}. The KIC is a photometric catalogue with estimated parameters of all stars
down to $V \simeq 18$ in the \kepler\ field of view. 
The values of \feh\ have been shown to be inaccurate \citep{molenda},
and we here confirm this based on the present analyses.




\begin{figure*}[t]
\centering
\includegraphics[width=16cm]{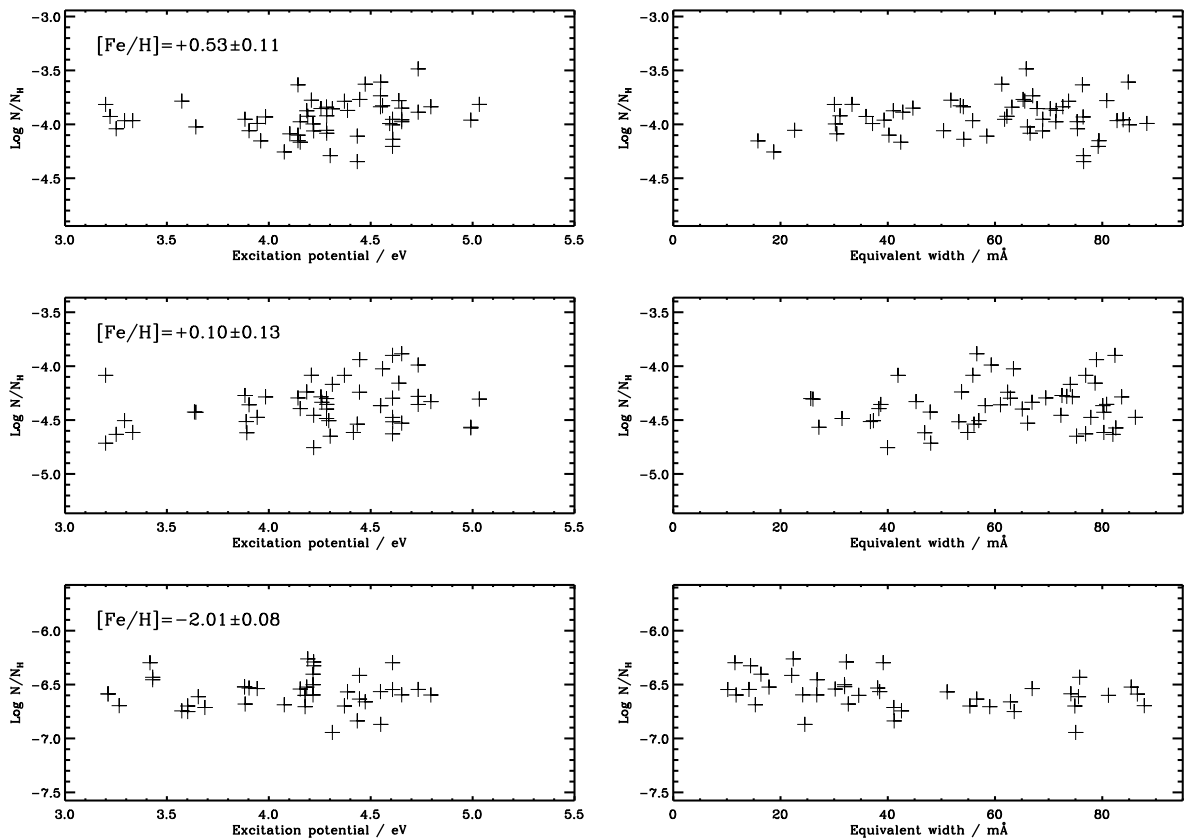}
\caption{Examples of diagnostic plots of \feone\ abundance vs.\ excitation potential and 
equivalent width for three different giants with KIC-IDs 11342694, 4157282, and 8017159 (top to bottom).}
\label{eqwexi}
\end{figure*}

\section{Target selection}



We have selected a sample of stars with a range of luminosities and metallicities, according to the KIC. Our target stars are in quite advanced stages
of evolution and the effects of the assumed physics in the theoretical models will be pronounced.
The lifetimes of the modes are long enough to permit 
asteroseismic extraction of information from individual modes.
The amplitudes of the modes and the large and small frequency separations for stars
on the giant branch have already been measured \citep{bedding}. However,
more sophisticated diagnostics will become available, when the duration of the time series is extended.

In the solar-metallicity open cluster M67 \citep{stello1} oscillations were shown to be present
in a few cases, while no evidence for oscillations were seen in metal-poor globular cluster M4
\citep{frandsen}. It is indeed possible that low metallicity may lead to smaller pulsation
amplitudes \citep{stello2}. In order to verify this hypothesis and to explore the
metal dependency of the oscillations, we have chosen a set of targets with a wide spread in
metallicity. The metal poor stars are at the same time among the oldest stars in the Galaxy. The
determination of accurate ages from the asteroseismic analysis is therefore of special interest. 
The KIC contains a small number of possibly 
metal poor K~giants, but as mentioned 
the metallicity from the catalogue is very uncertain.



To verify that our adopted technique is valid, 
we analysed 6 bright K~giant targets. 
They were selected from the work of \cite{smith} and from 
the PASTEL catalogue \citep{soubiran} with 
the criterion that they have accurate values and 
represent a relatively wide range in the atmospheric parameters.

\begin{table*}
\caption{Atmospheric parameters of the 14 \kepler\ K~giant targets as determined from VWA.
The \logg\ value is determined from the \feone/\fetwo\ ionization balance and the
wide Ca lines at $\lambda6122$ and $\lambda6162$\,\AA. 
These measures are combined and the weighted mean is given as $\left\langle \mathrm{log} \, g\right\rangle$.}
\centering
\begin{tabular}{rr@{}lcccr}
\hline \hline 
KIC-ID   & \multicolumn{2}{c}{\teff} & \logg\ (\feone/\fetwo) & \logg\ (Ca $\lambda$6122) & \logg\ (Ca $\lambda$6162) & $\left\langle \mathrm{log} \, g\right\rangle$ \\
\hline
 1726211 & $4950$&$\pm70$  & $2.29\pm0.10$ & --            & $2.80\pm0.26$ & $ 2.36\pm0.26$ \\
 2714397 & $5000$&$\pm70$  & $2.68\pm0.08$ & $2.23\pm0.44$ & $2.48\pm0.31$ & $ 2.65\pm0.25$ \\
 3744043 & $5020$&$\pm70$  & $3.06\pm0.07$ & $3.16\pm0.18$ & $3.10\pm0.15$ & $ 3.08\pm0.25$ \\
 3860139 & $4550$&$\pm90$  & $2.61\pm0.20$ & $1.98\pm0.20$ & $2.20\pm0.09$ & $ 2.23\pm0.25$ \\
 3936921 & $4580$&$\pm90$  & $2.11\pm0.17$ & --            & $2.27\pm0.17$ & $ 2.19\pm0.27$ \\
 4157282 & $4450$&$\pm90$  & $1.88\pm0.26$ & --            & --            & $ 1.88\pm0.35$ \\ 
 4177025 & $4390$&$\pm90$  & $1.93\pm0.22$ & --            & $1.74\pm0.25$ & $ 1.85\pm0.29$ \\
 5709564 & $4775$&$\pm70$  & $2.48\pm0.09$ & $2.15\pm0.34$ & $2.47\pm0.11$ & $ 2.46\pm0.25$ \\
 7006979 & $4770$&$\pm70$  & $2.22\pm0.06$ & --            & $2.52\pm0.31$ & $ 2.23\pm0.25$ \\
 8017159 & $4625$&$\pm70$  & $1.11\pm0.08$ & $2.21\pm0.28$ & --            & $ 1.19\pm0.25$ \\
 8476245 & $4865$&$\pm70$  & $1.86\pm0.09$ & $2.24\pm0.21$ & --            & $ 1.92\pm0.25$ \\
10403036 & $4485$&$\pm70$  & $1.90\pm0.18$ & --            & $2.02\pm0.09$ & $ 2.00\pm0.25$ \\
10426854 & $4955$&$\pm80$  & $2.38\pm0.15$ & $2.46\pm0.47$ & $2.73\pm0.22$ & $ 2.49\pm0.27$ \\
11342694 & $4695$&$\pm100$ & $3.10\pm0.18$ & $2.52\pm0.19$ & $2.46\pm0.26$ & $ 2.75\pm0.27$ \\

\hline
KIC-ID   &  \multicolumn{2}{c}{\feh} & $\xi_t$ [km/s] & $v_{\rm macro}$ [km/s] & \vsini\ [km/s] & $v_{\rm rad}$ [km/s] \\
\hline
 1726211 & $-0.66$&$\pm0.08$ & $1.36\pm0.40$ & 4.0$\pm1.0$ & 0.5$\pm1.0$ & $-145.1\pm0.5$ \\
 2714397 & $-0.40$&$\pm0.08$ & $1.40\pm0.25$ & 3.5$\pm1.0$ & 1.5$\pm1.0$ & $-191.6\pm0.5$ \\
 3744043 & $-0.25$&$\pm0.08$ & $1.10\pm0.14$ & 1.5$\pm1.0$ & 3.0$\pm1.0$ & $ -55.3\pm0.5$ \\
 3860139 & $+0.25$&$\pm0.13$ & $1.35\pm0.50$ & 4.0$\pm1.0$ & 2.5$\pm1.0$ & $ -25.2\pm0.5$ \\
 3936921 & $+0.29$&$\pm0.10$ & $1.15\pm0.56$ & 5.0$\pm1.0$ & 2.0$\pm1.0$ & $ -48.6\pm0.5$ \\
 4157282 & $+0.10$&$\pm0.13$ & $0.97\pm0.21$ & 3.0$\pm1.0$ & 2.5$\pm1.0$ & $ -36.7\pm0.5$ \\
 4177025 & $-0.25$&$\pm0.11$ & $1.40\pm0.40$ & 2.0$\pm1.0$ & 2.0$\pm1.0$ & $-123.5\pm0.5$ \\
 5709564 & $-0.22$&$\pm0.08$ & $1.50\pm0.08$ & 4.3$\pm1.0$ & 1.0$\pm1.0$ & $-105.9\pm0.5$ \\
 7006979 & $-0.36$&$\pm0.08$ & $1.50\pm0.16$ & 4.0$\pm1.0$ & 1.2$\pm1.0$ & $ -57.4\pm0.5$ \\
 8017159 & $-2.01$&$\pm0.08$ & $1.70\pm0.30$ & 3.0$\pm1.0$ & 3.0$\pm1.0$ & $-376.0\pm0.5$ \\
 8476245 & $-1.33$&$\pm0.09$ & $1.85\pm0.28$ & 3.0$\pm1.0$ & 3.0$\pm1.0$ & $-130.4\pm0.5$ \\
10403036 & $-0.58$&$\pm0.09$ & $1.35\pm0.09$ & 2.0$\pm1.0$ & 4.5$\pm1.0$ & $-125.7\pm0.5$ \\
10426854 & $-0.31$&$\pm0.10$ & $1.45\pm0.23$ & 2.0$\pm1.0$ & 3.3$\pm1.0$ & $ -45.7\pm0.5$ \\
11342694 & $+0.53$&$\pm0.11$ & $0.93\pm0.26$ & 2.0$\pm1.0$ & 3.7$\pm1.0$ & $ -20.0\pm0.5$ \\
\hline
\hline
\end{tabular}
\label{resulttable}
\end{table*}

\begin{figure}[t]
\centering
\hskip -0.8cm \includegraphics[width=9.5cm]{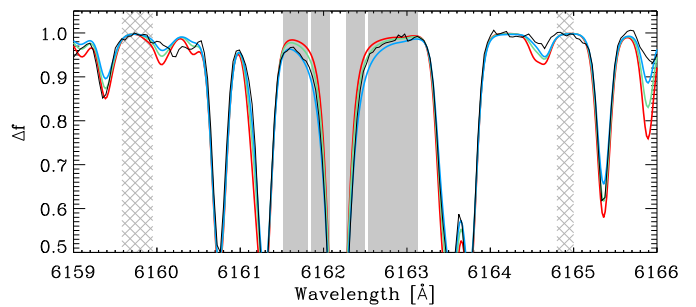}
\caption{Example of fitting the pressure-sensitive Ca line at 6162\AA$\ $in Arcturus ($\alpha$ Boo). Hatched regions are used to normalise the spectrum and $\chi^2$ is calculated in the four shaded regions. The best fit is the green line; the red/blue line has \logg\ lower/higher by 0.6 dex.}
\label{ca6162}
\end{figure}

\section{The observations}

We used spectra from a small pilot program carried out with the FIES spectrograph
at the NOT in 2008, followed by a larger project in 2009, where 5 nights were allocated. Unfortunately,
the number of spectra obtained in 2009 was small due to bad weather. In 2010 we have been much more
successful (7 nights allocated), and the data for 50 giant stars are now being processed.

The spectrograph was used in the high resolution mode ($R$ = 65,000) with ThAr calibration spectra framing
each target exposure. The exposure-meter\footnote{see http://www.not.iac.es/instruments/fies/} was used to get similar
signal-to-noise ratios (S/N) for all spectra aiming at 80--100. Exposure times varied from a few minutes for the
brighter giants to one hour for fainter targets split in two half hour exposures to reduce problems with cosmic rays.

The extraction was done with the software package FIEStool\footnote{see http://www.not.iac.es/instruments/fies/fiestool/FIEStool.html}
using the calibration frames recorded every night as a standard procedure at the NOT.

\section{Spectroscopic determination of atmospheric parameters}

The high-resolution spectra obtained were used to determine \teff, microturbulent velocity ($\xi_t$), \logg, \vsini\ and metallicity of the selected sample of giants. For the analysis, we used iron lines in the wavelength range 4500\AA$\ $to 7000\AA, avoiding the regions affected by telluric lines. The number of lines used depend on the S/N of the individual spectra and on the degree of blending between neighboring lines. As far as possible, non-blended lines were preferred, resulting in rather few lines ($\approx 40$) for some of the targets in the sample.  


\subsection{Atmospheric parameters from \feone/\fetwo-abundances}

We used the VWA software \citep{bruntt1,bruntt2,bruntt5,bruntt4} to determine the fundamental parameters of the targets. The software is a semi-automatic package in which a careful continuum normalization is done by manually selecting continuum points in the stellar spectrum by comparing it to a synthetic spectrum 
with similar fundamental parameters \citep{bruntt4}. 
This was followed by a careful selection of the least blended lines, each of which were iteratively fitted with a synthetic spectrum, including the contribution from weakly blending lines. This is important for the rich giant spectra, especially in the blue wavelength range. We adopted MARCS model atmospheres \citep{gustafsson} and atomic line data from the Vienna Atomic Line Database \citep{kupka}. Each line fit was inspected in great detail and bad fits discarded, 
resulting in between 40--120 \feone\ and 3--11 \fetwo\ lines 
that were used in the determination of the fundamental parameters. 
As initial guesses for the parameters of the model atmosphere, the values in the KIC were used. 
The $\xi_t$ and \teff\ parameters were then refined through several iterations to remove 
correlations between the abundances of \feone\ and equivalent width (EW) and excitation potential (EP), respectively. 
We also required agreement between the \feone\ and \fetwo\ abundances, which was established
by adjusting \teff\ and \logg. 
An example of the \feone\ abundances vs.\ EW and EP is shown in Fig.~\ref{eqwexi} for a giant  
with high, solar, and low metallicity. 
Note that abundances are measured relative to the same lines in 
the solar spectrum, as described by \cite{bruntt4}.

The results for the 14 giants are presented in Table \ref{resulttable}. 
All abundances are measured 
relative to the Sun, with the errors being the RMS scatter of the abundances of each line included in the fit. 
The uncertainties of \teff, \logg, and $\xi_t$ were calculated by changing one model parameter at a time, until at least a 3-$\sigma$ deviation was produced between the \feone/\fetwo\ abundances or on the slope of the abundances vs.\ EP or EW. 
From these, a 1-$\sigma$ error was calculated, giving the internal precision of the parameters calculated in VWA. 
To this we quadratically added a systematic error which we evaluate in Sect.~\ref{comp},
using the results for the six bright giants.

\begin{figure}
\centering
\includegraphics[width=8.7cm]{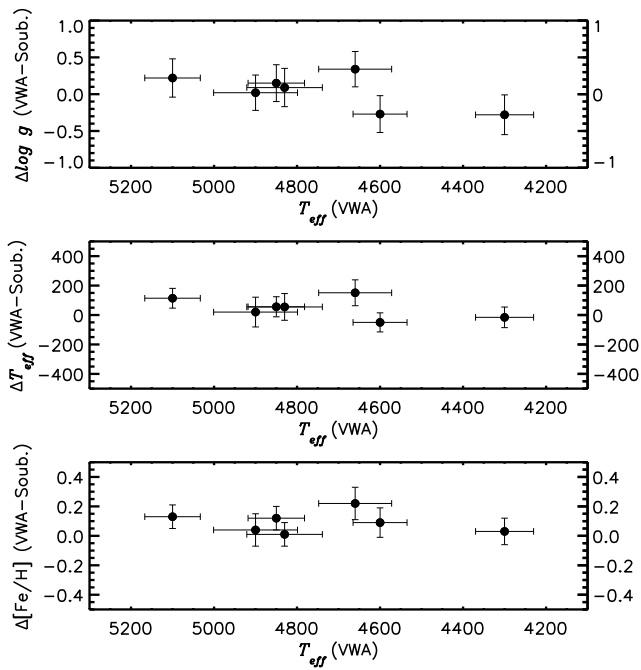}
\caption{Comparison of parameters determined from VWA and from the PASTEL catalogue \citep{soubiran} for six bright giants.}
\label{reference}
\end{figure}

\subsection{Log~$g$ from pressure sensitive lines}
It is possible, especially for cool stars, to determine \logg\ from a selection of strong, pressure-sensitive lines, thus putting further constraints on the surface gravity. 
Commonly used lines are the Mg\,{\sc i}b, Na\,{\sc i}~D, and the Ca\,{\sc i} lines at 6122\AA$\ $and 6162\AA. Determination of \logg\ for giants from the Mg\,{\sc i}b and
Na\,{\sc i}~D lines was problematic since the degree of line blending or the S/N in these areas did not enable us to make a trustworthy continuum determination. 
Also the Mg\,{\sc i}b lines are so wide and lying so close that there is no continuum present between them, 
making the normalization even more difficult. 
Furthermore, we found that these lines are quite insensitive to changes in the surface gravity. We therefore only used the Ca lines for the \logg\ determination. In some cases the degree of line blending 
around one of the Ca lines was so severe that a reliable fit could not be made, thus only one Ca line was used to estimate the \logg\ value. For a single target (KIC 4157282) this was 
the case for both Ca lines, so only the matching \feone/\fetwo\ abundances could be used. 

An example of fitting the Ca 6162 line is shown in Fig.~\ref{ca6162}.
The observed spectrum around the Ca lines was fitted with three synthetic spectra, each with a different value of \logg. The $\chi^2$ value was then calculated for each fit to determine the best value for \logg. The method is described in greater detail by \citet{bruntt4}. The best fitting values for \logg\ are quoted in Table \ref{resulttable}, where the weighted mean of the determination from the \feone/\fetwo\ abundances and from the Ca lines is taken as the final result,
and given in the last column.

\subsection{Determination of \vsini\ and macroturbulence}
To estimate the values of \vsini\  and macroturbulence, we evaluated by visual inspection the fits of synthetic line profiles to dozens of isolated lines throughout spectrum. The results quoted in Table \ref{resulttable} were taken as the average of the values found from the individual fits. It is seen that both the macroturbulence and \vsini\  for the giants is more or less the same and the surface rotation is very slow as expected for giant stars. The uncertainty of the parameters was estimated by changing the parameters until a significant deviation from the observed profile was seen in the fit. The deviation was found by visual inspection of the fitted line profiles and is thus only a rough estimate of the uncertainty.

\subsection{Discussion}\label{comp}

To validate our method we obtained FIES spectra of six bright giants, 
chosen from the works of \citet{soubiran} and \citet{smith}.
\citet{soubiran} used the TGMET method for their analysis, 
where a large grid of spectra from the ELODIE spectrograph are assigned parameters from an extensive literature search. 
The parameters of each star are then found by locating the best matching spectrum in the library. We compare the results in
Fig.~\ref{reference} and Table~\ref{soubtable}. 
There are no significant 
offsets for \logg\ or \teff\ between the TGMET method and VWA, although our \feh\ values are slightly
higher by about 0.1~dex. The RMS scatter of the differences is 60K for \teff, 0.24 dex for \logg\ and 0.07 dex for \feh.
We adopt these as the ``systematic errors'' that are due to the differences in the adopted method, grid of model atmospheres,
spectrum normalization etc. We are aware that a rigorous treatment of each of these effects would require
a much larger sample of stars, and is beyond the scope of this work.

\begin{table*}
\caption{Comparison of the spectroscopic parameters from VWA 
with photometric values from KIC and from the asteroseismic analysis for 14 \kepler\ giants.}
\centering
\begin{tabular}{r|r@{}lcc|ccc|r@{}lc}
\hline \hline

    & \multicolumn{4}{c|}{VWA}     & \multicolumn{3}{c|}{KIC} & \multicolumn{3}{c}{Asteroseis.} \\
KIC-ID & \multicolumn{2}{c}{\teff} & \logg & \feh & \teff & \logg & \feh & \multicolumn{2}{c}{\teff} & \logg \\ \hline

   1726211 & $4950$&$\pm70$  & $ 2.36\pm0.26$ & $ -0.66\pm0.08$ & $ 4837\pm200$ & $2.68\pm0.50$ & $ -0.96\pm0.50$ & $ 4627$&$\pm  103$ & $2.37\pm0.04$ \\
   2714397 & $5000$&$\pm70$  & $ 2.65\pm0.25$ & $ -0.40\pm0.08$ & $ 4881\pm200$ & $2.52\pm0.50$ & $ -0.53\pm0.50$ & $ 4762$&$\pm   99$ & $2.41\pm0.01$ \\
   3744043 & $5020$&$\pm70$  & $ 3.08\pm0.25$ & $ -0.25\pm0.08$ & $ 4994\pm200$ & $2.50\pm0.50$ & $ -0.09\pm0.50$ & $ 4769$&$\pm  115$ & $2.95\pm0.02$ \\
   3860139 & $4550$&$\pm90$  & $ 2.23\pm0.25$ & $ +0.25\pm0.13$ & $ 4589\pm200$ & $2.22\pm0.50$ & $ +0.60\pm0.50$ &      $$&$$         &               \\
   3936921 & $4580$&$\pm90$  & $ 2.19\pm0.27$ & $ +0.29\pm0.10$ & $ 4436\pm200$ & $2.38\pm0.50$ & $ -0.06\pm0.50$ & $ 4587$&$\pm   54$ & $2.33\pm0.04$ \\
   4157282 & $4450$&$\pm90$  & $ 1.88\pm0.35$ & $ +0.10\pm0.13$ & $ 4344\pm200$ & $2.13\pm0.50$ & $ -0.78\pm0.50$ &      $$&$$         &               \\   
   4177025 & $4390$&$\pm90$  & $ 1.85\pm0.29$ & $ -0.25\pm0.11$ & $ 4346\pm200$ & $2.14\pm0.50$ & $ -0.49\pm0.50$ &      $$&$$         &               \\   
   5709564 & $4775$&$\pm70$  & $ 2.46\pm0.25$ & $ -0.22\pm0.08$ & $ 4752\pm200$ & $2.52\pm0.50$ & $ -0.06\pm0.50$ & $ 4718$&$\pm  118$ & $2.33\pm0.04$ \\
   7006979 & $4770$&$\pm70$  & $ 2.23\pm0.25$ & $ -0.36\pm0.08$ & $ 4891\pm200$ & $2.21\pm0.50$ & $ -0.01\pm0.50$ & $ 4645$&$\pm  103$ & $2.45\pm0.05$ \\
   8017159 & $4625$&$\pm70$  & $ 1.19\pm0.25$ & $ -2.01\pm0.08$ & $ 4634\pm200$ & $2.45\pm0.50$ & $ -1.07\pm0.50$ &      $$&$$         &               \\   
   8476245 & $4865$&$\pm70$  & $ 1.92\pm0.25$ & $ -1.33\pm0.09$ & $ 4817\pm200$ & $2.76\pm0.50$ & $ -1.20\pm0.50$ &      $$&$$         &               \\   
  10403036 & $4485$&$\pm70$  & $ 2.00\pm0.25$ & $ -0.58\pm0.09$ & $ 4388\pm200$ & $2.21\pm0.50$ & $ -1.39\pm0.50$ &      $$&$$         &               \\   
  10426854 & $4955$&$\pm80$  & $ 2.49\pm0.27$ & $ -0.31\pm0.10$ & $ 4731\pm200$ & $2.57\pm0.50$ & $ -1.03\pm0.50$ &      $$&$$         &               \\   
  11342694 & $4695$&$\pm100$ & $ 2.75\pm0.27$ & $ +0.53\pm0.11$ & $ 4603\pm200$ & $2.65\pm0.50$ & $ +0.50\pm0.50$ & $ 4670$&$\pm    90$ & $2.78\pm0.02$  \\
 
\hline
\hline
\end{tabular}
\label{comptable}
\end{table*}

\begin{figure}[t]%
\centering
\includegraphics[width=8.7cm]{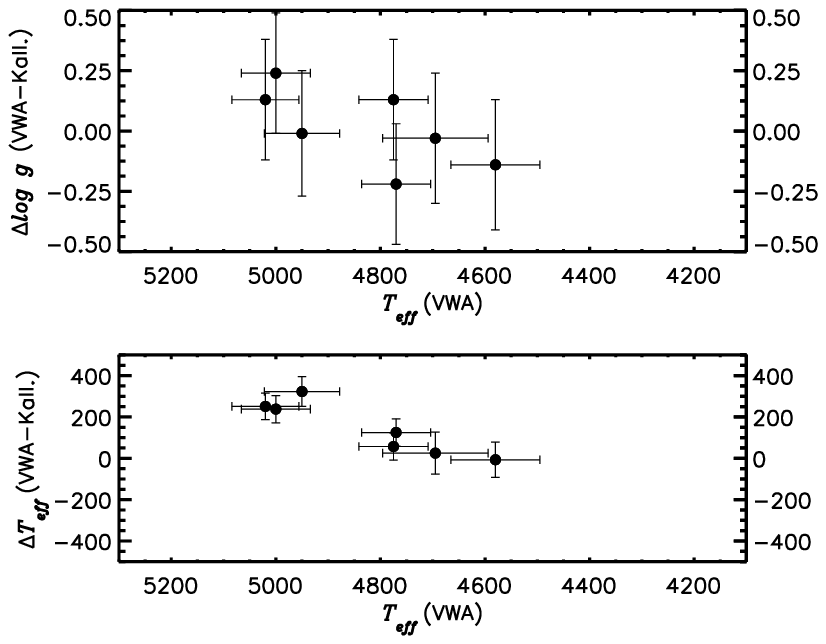}
\caption{Comparison of \logg\ and \teff\ determined 
from VWA and the asteroseismic method of \cite{kallinger10}.} 
\label{kal}%
\end{figure}

\begin{figure*}[t]
\centering
\includegraphics[width=8.5cm]{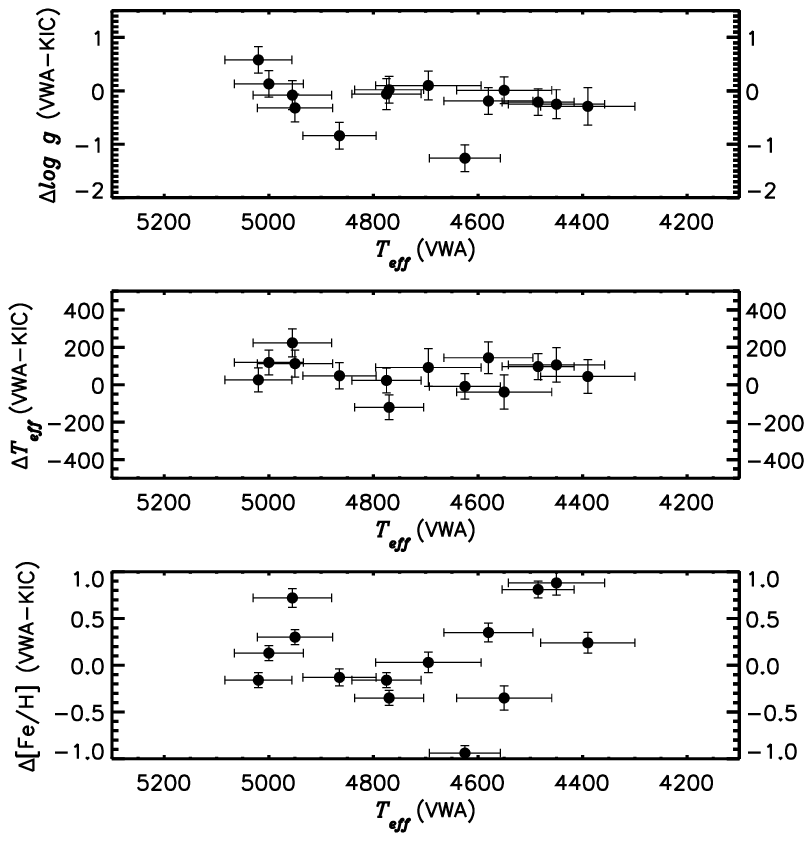}
\hskip 0.6cm
\includegraphics[width=8.5cm]{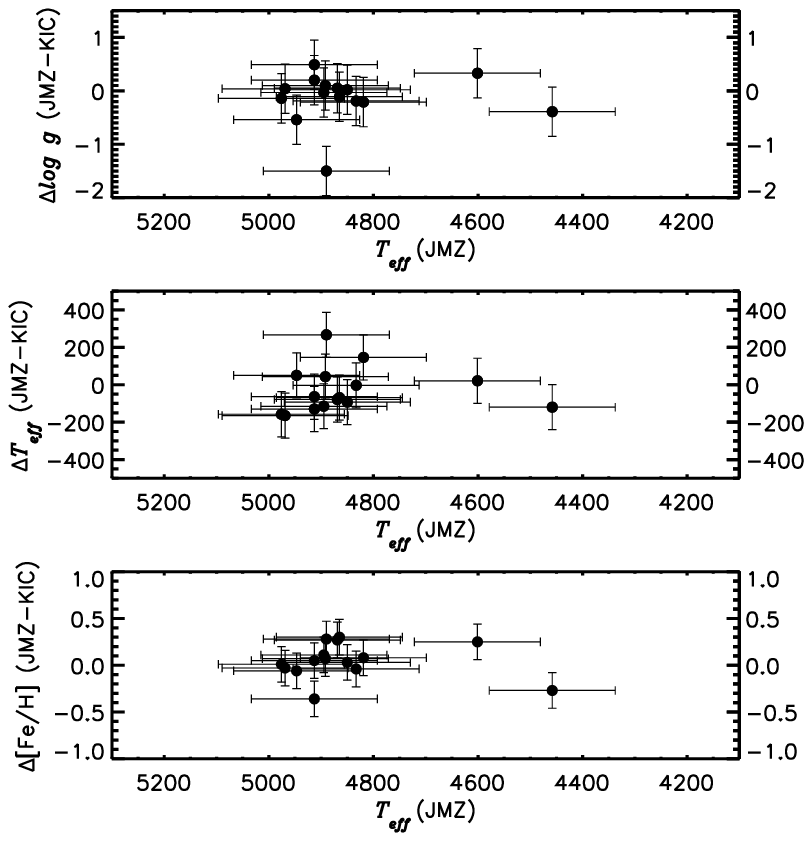}
\caption{Left panels: Comparison of atmospheric parameters determined with VWA and from KIC.
Right panels: Comparison of atmospheric parameters determined with ROTFIT \citep{molenda} and KIC.}
\label{kepgig}
\end{figure*}


An analysis of the \kepler\ light curves has already been carried out for seven 
of the targets in our sample \citep{kallinger10}.
This analysis gives values of \teff\ and \logg\ and relies on the comparison of the asteroseismic 
data with an evolutionary model grid. As seen in Fig.~\ref{kal} there is good agreement between the \logg\ values found 
from VWA and the values from the asteroseismic analysis. 
However, the comparison of \teff\ in the lower panel shows a correlation with \teff, 
which may indicate a systematic problem in the approach of \cite{kallinger10} for the hottest giants. 
The mean offsets and RMS values are $\Delta$\logg=$0.01\pm0.27$\,dex and $\Delta$\teff=$145\pm127$\,K, respectively. 
The asteroseismic values are tabulated in Table \ref{comptable} along with the parameters from the KIC.

In Fig.~\ref{kepgig} and Table \ref{comptable} we compare the atmospheric parameters from KIC 
with the VWA analysis (left panels) and the spectroscopic study of \cite{molenda} (right panels). 
The two studies show the same overall picture, 
but our sample includes significantly more of the evolved stars with $T_{\rm eff} < 4800$\,K.
We will now discuss the comparison of VWA and the KIC values. 
It is seen that there is large scatter when comparing \feh\ from VWA and KIC: 
the range is from $-1$ to $+1$ dex with the average difference and RMS scatter 
being $\Delta {\rm [Fe/H]} = +0.10\pm0.50$. 
For \logg\ we notice that the two stars with the largest discrepancy
also have the lowest metallicity (KIC-IDs 8017159 and 8476245), which may indicate
a problem in KIC for the low-metallicity stars.
If we consider only the remaining one dozen stars we obtain $\Delta \log g = -0.05\pm0.25$, 
which indicates that the KIC values are fairly robust.
Finally, the spectroscopic effective temperatures are in fairly good agreement 
with the KIC values, $\Delta T_{\rm eff} = +62\pm85$\,K.

From our sample of 14 Kepler giants, we conclude that the KIC values for \teff\ and \logg\ are 
trustworthy for target selection and statistical studies of their asteroseismic properties, 
but only for ${\rm [Fe/H]} > -0.5$ dex.
It is clear that for detailed asteroseismic analyses we need homogeneously
determined spectroscopic parameters.



%

\section{Conclusion}

We have determined accurate atmospheric parameters for a sample of 14 K~giant 
targets that are being observed with the NASA \kepler\ satellite. 
The parameters are mandatory to put constraints on asteroseismic models when comparing observations and theory. 
We confirm the results by \cite{molenda} that there are serious discrepancies for \feh\ when comparing 
with the photometric KIC catalogue (RMS scatter in \feh\ of 0.5 dex), 
while \teff\ and \logg\ values are in reasonable agreement.
However, for \logg\ we find discrepancies of about $1$ dex for two stars with ${\rm [Fe/H]} < -1.0$,
indicating that there may be a problem in the KIC catalogue at low metallicities.

We have validated our method and evaluated systematic errors from the analysis of 6 bright giants 
with well-known parameters and compared with results in the literature, 
confirming that our analysis is reliable. 
Also we see good agreement between our parameters and the ones found from asteroseismology.

The uncertainties on \logg\ and \feh\ in KIC 
are too large to match the quality of the data produced by \kepler, and this emphasizes 
the importance and need for further, detailed spectroscopic studies of the \kepler\ giant targets. 
This paper will be followed by a second paper presenting the results for an additional 50 K~giants.

We have verified that one of the \kepler\ giants is a population II star (KIC 8017159), and we expect to find
several more in our larger sample of stars. Until now, only one nearby population~II star, $\nu$~Ind, 
has been studied using asteroseismic techniques \citep{nuind}.

%

\begin{table}
\caption{Comparison of parameters from VWA and the PASTEL catalogue \citep{soubiran} for six bright giants.}
\centering
\setlength{\tabcolsep}{3pt} 
\begin{tabular}{r|ccc|ccc}
\hline \hline

    & 
\multicolumn{3}{c|}{VWA} & \multicolumn{3}{c}{PASTEL}\\
ID & \teff & \logg & \feh & \teff & \logg & \feh       \\ \hline
      
 $\alpha$~Mon  & $4850\pm70$ & $ 2.77\pm0.25$ & $ +0.08\pm0.08$ & $ 4794 $ & $ 2.62$ & $ -0.04 $ \\
 $\mu$~Leo     & $4660\pm90$ & $ 2.63\pm0.24$ & $ +0.53\pm0.11$ & $ 4509 $ & $ 2.29$ & $ +0.31 $ \\
 $\alpha$~Boo  & $4300\pm70$ & $ 1.43\pm0.27$ & $ -0.52\pm0.09$ & $ 4316 $ & $ 1.71$ & $ -0.55 $ \\
 $\lambda$~Peg & $4830\pm90$ & $ 2.56\pm0.26$ & $ -0.08\pm0.08$ & $ 4775 $ & $ 2.47$ & $ -0.09 $ \\
 $\mu$~Peg     & $5100\pm70$ & $ 2.96\pm0.26$ & $ +0.05\pm0.08$ & $ 4986 $ & $ 2.74$ & $ -0.08 $ \\
 $\psi$~UMa    & $4600\pm70$ & $ 2.11\pm0.25$ & $ -0.04\pm0.10$ & $ 4605 $ & $ 2.38$ & $ -0.13 $ \\ 

\hline
\hline

\multicolumn{7}{l}{Uncertainties on \teff, \logg, and \feh\ for PASTEL are 80\,K, 0.1 dex}\\
\multicolumn{7}{l}{and 0.1 dex, respectively.}

\end{tabular}
\label{soubtable}
\end{table}

\begin{acknowledgements}
We are grateful for the many useful comments from Dennis Stello and to
Thomas Kallinger for sharing results from his asteroseismic analyses.
This research took advantage of the SIMBAD and VIZIER databases at the CDS, Strasbourg 
(France), and NASA's Astrophysics Data System Bibliographic Services.
\end{acknowledgements}

\bibliographystyle{aa}
\bibliography{bruntt-kepler-giants-01} 

\end{document}